\documentclass[twocolumn,conference]{IEEEtran}
\usepackage[T1]{fontenc}
\usepackage[latin9]{inputenc}
\usepackage{array}
\usepackage{multirow}
\usepackage{amsmath}
\usepackage{amssymb}
\usepackage{graphicx}
\usepackage[unicode=true,
 bookmarks=true,bookmarksnumbered=true,bookmarksopen=true,bookmarksopenlevel=1,
 breaklinks=false,pdfborder={0 0 0},pdfborderstyle={},backref=false,colorlinks=false]
 {hyperref}
\hypersetup{pdftitle={Your Title},
 pdfauthor={Your Name},
 pdfpagelayout=OneColumn, pdfnewwindow=true, pdfstartview=XYZ, plainpages=false}

\makeatletter

\providecommand{\tabularnewline}{\\}

\usepackage[caption=false,font=footnotesize]{subfig}

\usepackage{cite}
\usepackage{algorithm}
\usepackage{algpseudocode}

\usepackage{hyperref}
\hypersetup{draft}
\usepackage{enumerate}
\usepackage[shortlabels]{enumitem}

\makeatother

\begin{document}
\title{Heterogeneously-Distributed Joint Radar Communications: Bayesian Resource
Allocation}
\author{\IEEEauthorblockN{Linlong~Wu\IEEEauthorrefmark{1}, Kumar~Vijay~Mishra\IEEEauthorrefmark{1},
Bhavani~Shankar~M.~R.\IEEEauthorrefmark{1} and Bj\"{o}rn~Ottersten\IEEEauthorrefmark{1}}\IEEEauthorblockA{\IEEEauthorrefmark{1}Interdisciplinary Centre for Security, Reliability
and Trust (SnT), University of Luxembourg\\
Email: \{linlong.wu@, kumar-mishra@ext., bhavani.shankar@, bjorn.ottersten@\}uni.lu}}
\maketitle
\begin{abstract}
Due to spectrum scarcity, the coexistence of radar and wireless communication
has gained substantial research interest recently. Among many scenarios,
the heterogeneously-distributed joint radar-communication system is
promising due to its flexibility and compatibility of existing architectures.
In this paper, we focus on a heterogeneous radar and communication
network (HRCN), which consists of various generic radars for multiple
target tracking (MTT) and wireless communications for multiple users.
We aim to improve the MTT performance and maintain good throughput
levels for communication users by a well-designed resource allocation.
The problem is formulated as a Bayesian Cram\'{e}r-Rao bound (CRB)
based minimization subjecting to resource budgets and throughput constraints.
The formulated nonconvex problem is solved based on an alternating
descent-ascent approach. Numerical results demonstrate the efficacy
of the proposed allocation scheme for this heterogeneous network. 
\end{abstract}


\IEEEpeerreviewmaketitle{}

\section{Introduction}

Although wireless communications and radar systems have developed
in parallel for decades, both share many aspects essentially in terms
of signal processing algorithms, devices and even architectures \cite{zhang2021overview}.
These common grounds, together with spectrum scarcity, have recently
motivated substantial research interest in the co-design of the integrated
sensing and communications \cite{liu2020joint}. In general, it can
be categorized into two research directions. The first direction aims
to develop a dual-functional system which simultaneously performs
radar and communication functionalities \cite{hassanien2019dual},
while the second one focuses on the coexistence of two separated radar
and communication systems \cite{zheng2019radar}. For both directions,
a well-designed resource allocation (RA) is desired (i.e., transmit
power, dwell time, spectrum, etc.) to avoid significant degenerated
performance and fully release the potential of this integrated network.

Many works have contributed to the RA of the integrated network. In
\cite{li2016joint}, radar waveform and adaptive communication transmission
scheme were designed to maximize signal-to-interference-plus-noise
ratio (SINR) while ensuring the communication system meeting certain
rate and power constraints. The work \cite{shi2016low} proposed to
minimize the total noise jamming power by optimizing the multicarrier
jamming power allocation. By leveraging the multicarrier waveforms,
\cite{wang2019power} proposed two co-design paradigms to improve
the spectrum efficiency by optimizing the power allocation and the
communication throughput. In \cite{zheng2017joint}, the compound
rate was proposed as the optimization metric, based on which the optimum
transmit policies for the coexistence was derived. The RA strategy
of distributed radar architecture for the target localization was
first analyzed in \cite{godrich2011power}. Subsequently in \cite{yan2020optimal},
asynchronous RA schemes were proposed for the heterogeneous radar
architecture, which was further extended in \cite{yan2021target}
to maximize the number of the targets that can be tracked.

\begin{figure}[tbh]
\begin{centering}
\includegraphics[scale=0.35]{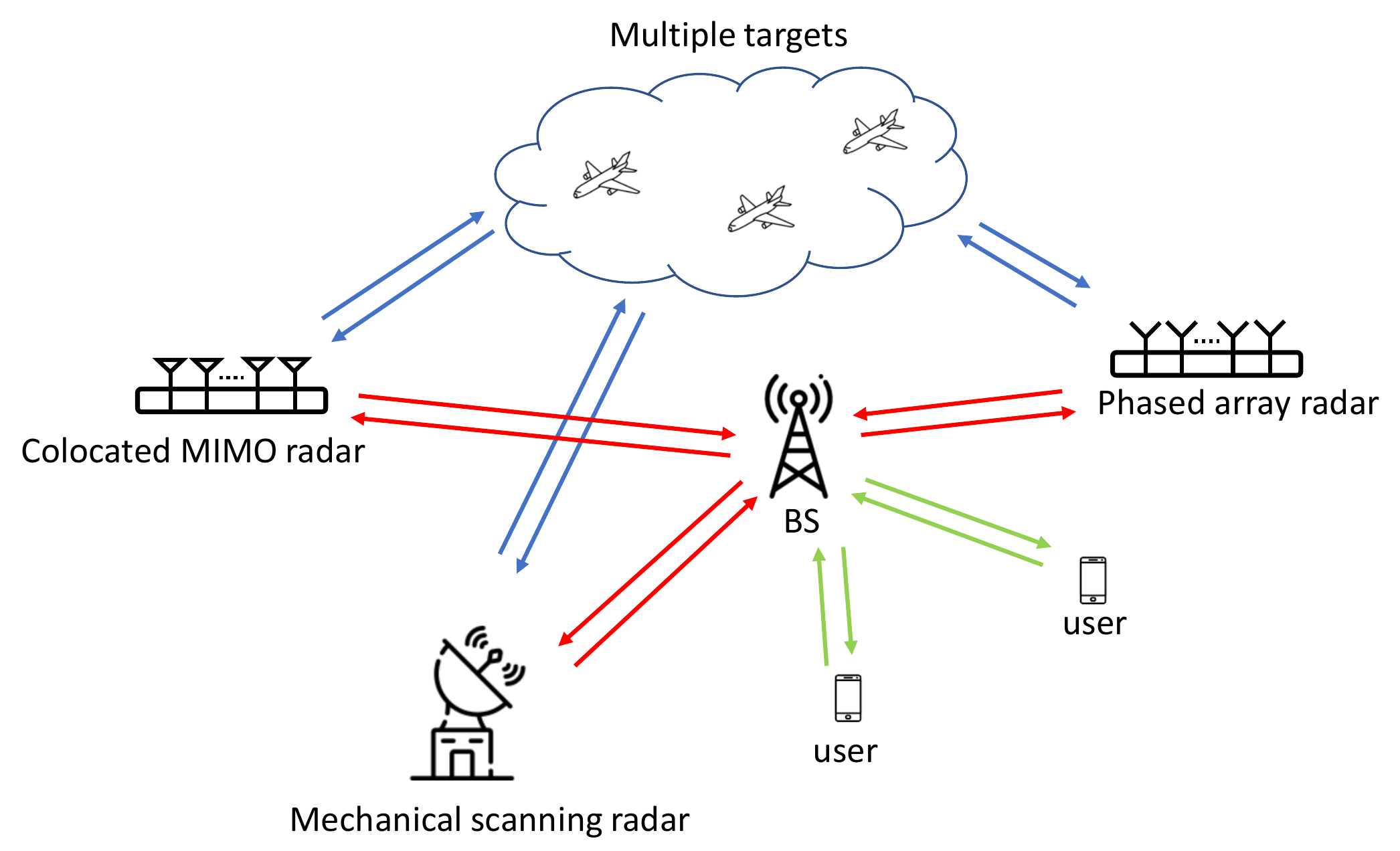}
\par\end{centering}
\caption{\label{fig:Sketch-of-the}Sketch of the considered scenario.}
\end{figure}

However, the existing works have not considered the heterogeneity
inside this integrated network. Throughout this paper, we focus on
a heterogeneous radar and communication network (HRCN), which consists
of different generic radars for tracking multiple targets and wireless
communications for multiple users. the sketch of the HRCN is shown
in Figure \ref{fig:Sketch-of-the}. Through resource allocation, we
aim to optimize the multiple target tracking (MTT) performance measured
by a Bayesian CRB based metric subject to meeting the communication
throughput requirements and the resource budgets. The formulated problem
is then recast into a maximin problem, to which we propose the alternating
descent-ascent method (ADAM). Numerical results are presented to demonstrate
the efficacy of the design allocation for this heterogeneous network.
To the best of our knowledge, this is the first paper dealing with
the heterogeneous resource allocation for integrated radar and communication
network. 

\section{HRCN Configurations and Target Tracking Model}

\subsection{Configurations of the HRCN}

\paragraph{Configuration of radars}

The $N$ radars consist of $N_{cr}$ co-located MIMO radars (MMRs),
$N_{pr}$ phased array radars (PARs) and $N_{mr}$ mechanical scanning
radars (MSRs) with $N=N_{cr}+N_{pr}+N_{mr}$. These radars are located
at the coordinates $\left\{ \left(x_{i},y_{i}\right)\right\} _{i=1}^{N}$.
The $N$ radars aim to track $Q$ widely separated point-like targets.
For each type of radar, we have the following assumptions:

\begin{enumerate}[start=1,label={(R\arabic*)}]
\item Colocated MMR, denoted by $i\in \varphi_{c}\triangleq\left\{ 1,\ldots,N_{cr}\right\} $, adopts the multiple beams to illuminates multiple targets simultaneously with the same dwell time length but different transmit powers. Hence, the revisit time intervals for all targets are the same. 
\item PAR, denoted by $i\in\varphi_{p}\triangleq\left\{ N_{cr}+1,\ldots,N_{cr}+N_{pr}\right\} $, adaptively rotates the beam to illuminate multiple targets sequentially, with the same transmit power but different dwell time lengths. Hence, the revisit time intervals for the different targets may not be the same.
\item MSR, dednoted by $i\in\varphi_{m}\triangleq\left\{ N_{cr}+N_{pr}+1,\ldots,N\right\} $, rotates mechanically and illuminates all the target sequentially with the same power and dwell time. Hence, the revisit time intervals for all targets are the same.
\item Mutual interference among the $N$ radars is assumed negligible due to the directional beams.
\end{enumerate}

\begin{figure*}[t]
\begin{centering}
\includegraphics[scale=0.39]{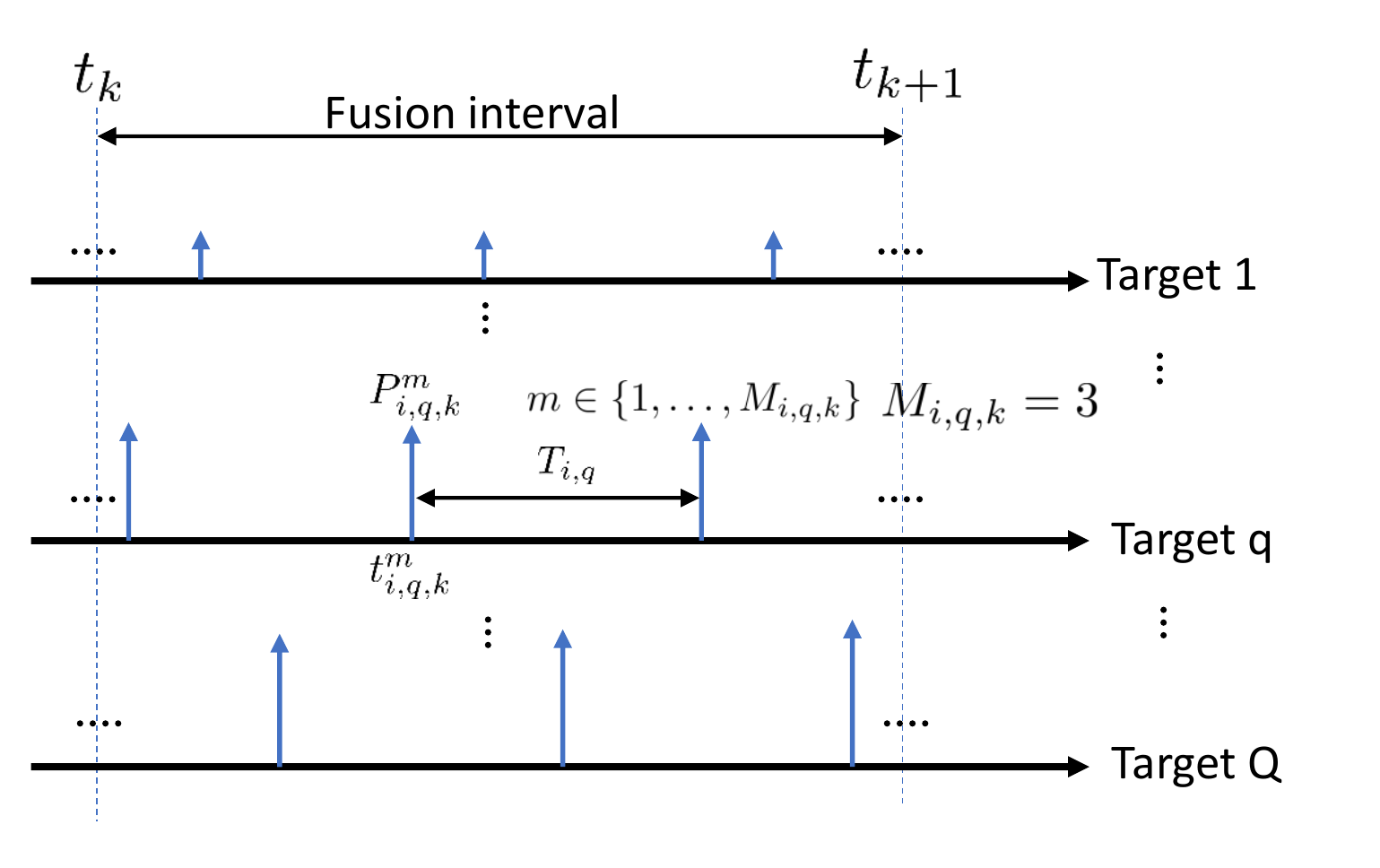}\includegraphics[scale=0.39]{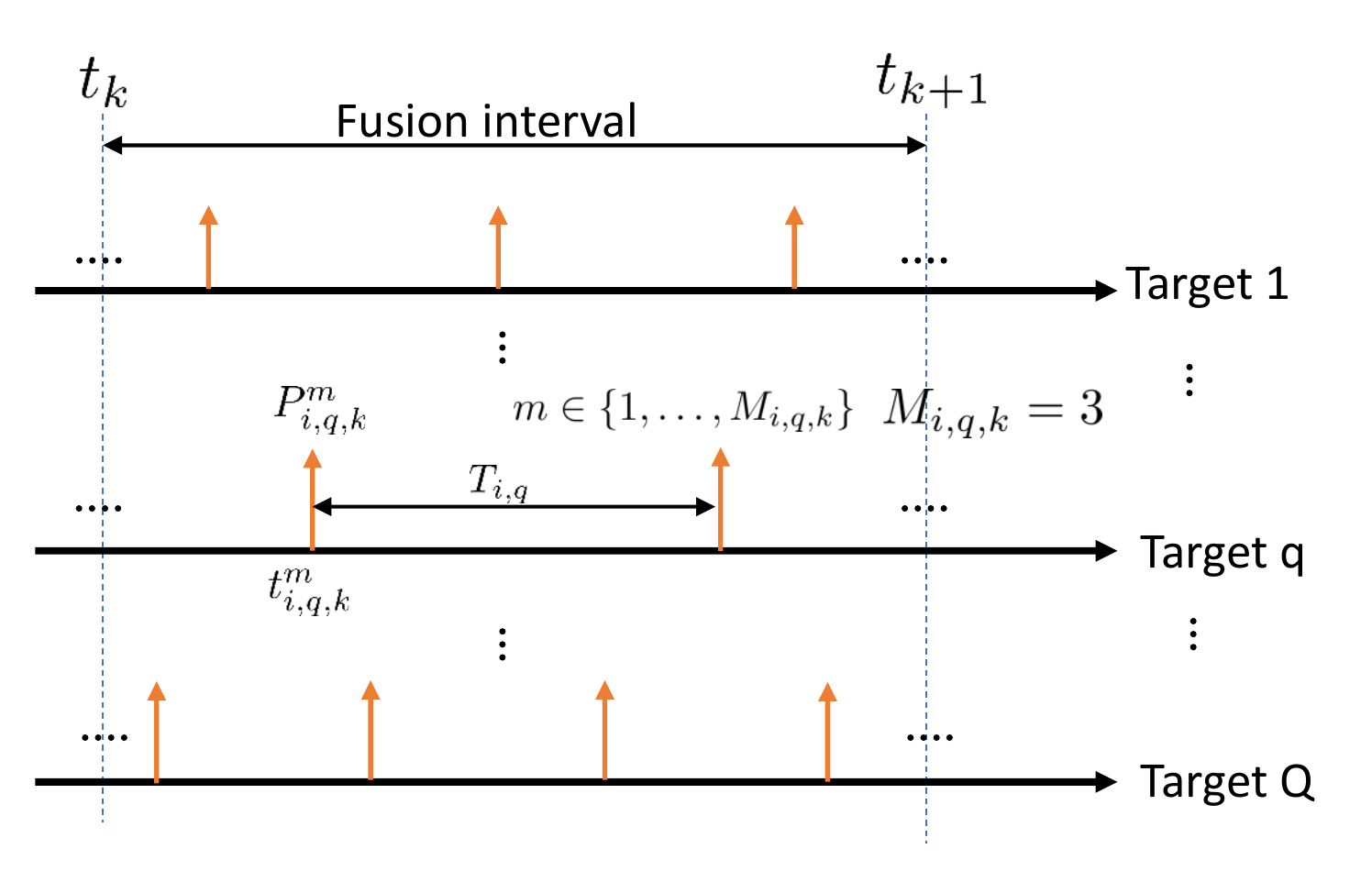}\includegraphics[scale=0.39]{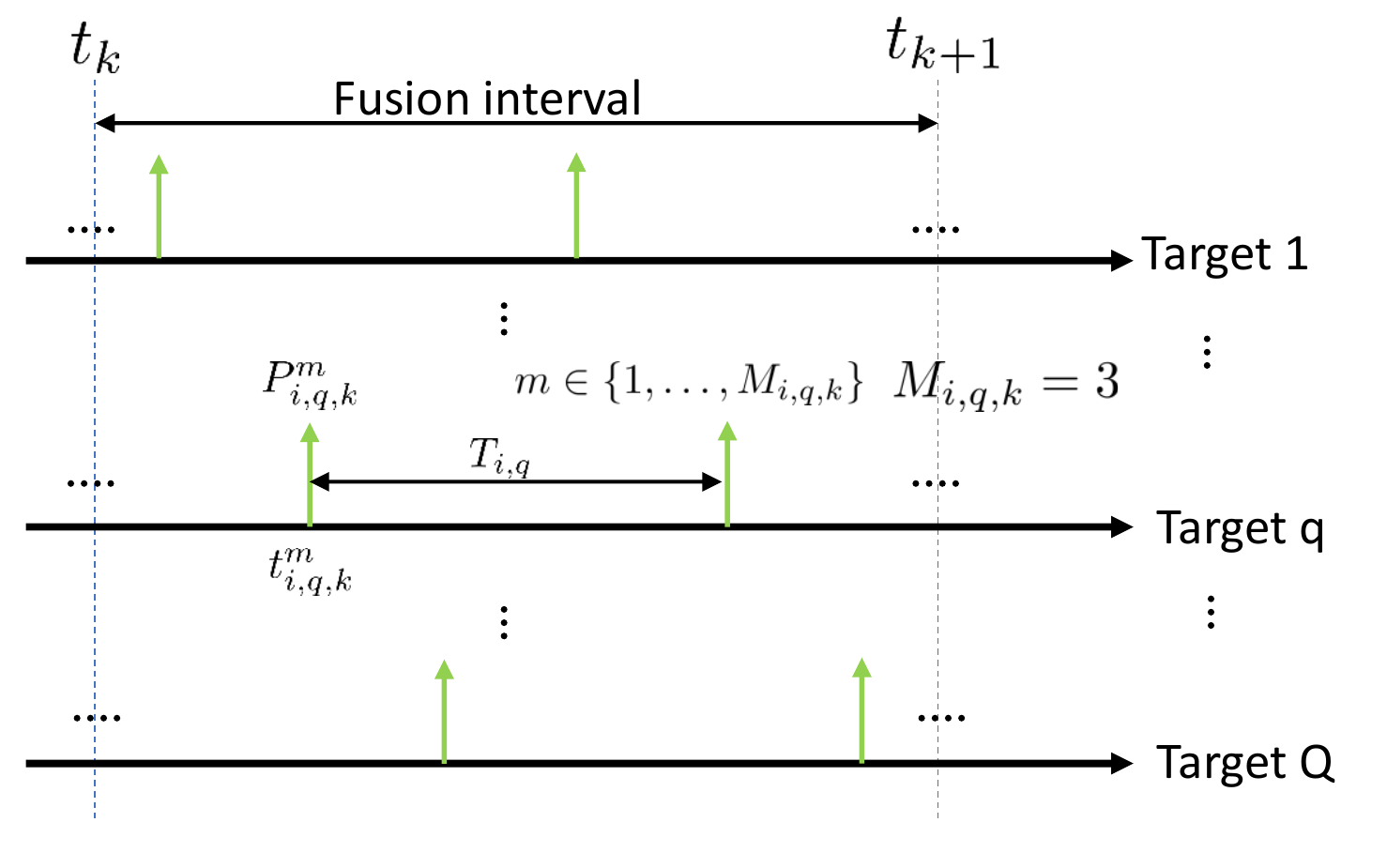}
\par\end{centering}
\caption{\label{fig:Operation-schemes-for}Operation schemes for radars. Left:
Colocated MIMO radar; middle: phased array radar; right: mechanical
scanning radar.}
\end{figure*}

The radar operation schemes are shown in Figure \ref{fig:Operation-schemes-for},
where $M_{i,q,k}$ is the number of measurements for target $q$ by
radar $i$ during the $k$-th fusion interval with the length $T_{0}=t_{k+1}-t_{k}$,
$t_{i,q,k}^{m},\forall m=1,\ldots,M_{i,q,k}$ is the $m$-th measurement
time, and $P_{i,q,k}^{m}$ is the transmit power corresponding to
measurements at $t_{i,q,k}^{m}$.

\paragraph{Configuration of wireless communications }

In the cellular communication system, $J$ downlink users are served
by a base station with the following configuration:

\begin{enumerate}[start=1,label={(C\arabic*)}]
\item All $J$ users are operating in different frequencies, i.e., no mutual interference among them.
\item During the $k$-th fusion time of the radar networks, the $j$-th downlink lasts with a constant power $P_{c,k}^{j}$. 
\item Gaussian codebook based transmission is assumed. 
\end{enumerate}

\paragraph{Interference between radars and communications}

During a fusion interval, the following assumptions holds for the
mutual interference:

\begin{enumerate}[start=1,label={(I\arabic*)}]
\item The $J$ users are affected by the radar signals intermittently. The complex parameter $\alpha_{i,j}^{c}$ accounts for the interference from the $j$-th downlink to the $i$-th radar. 
\item The $N$ radars are interfered persistently by the $J$ downlinks. The complex parameter $\alpha_{j,i}^{r}$ accounts for the interference from the $i$-th radar to the $j$-th downlink.
\end{enumerate}

\subsection{Target Tracking Model}

In the $k$-th fusion period, the state of target $q$ at the receiving
time $t_{i,q,k}^{m}$ is defined as 
\begin{equation}
\boldsymbol{s}_{i,q,k}^{m}=\left[x_{i,q,k}^{m},\dot{x}_{i,q,k}^{m},y_{i,q,k}^{m},\dot{y}_{i,q,k}^{m}\right]^{T},
\end{equation}
where $\left(x_{i,q,k}^{m},y_{i,q,k}^{m}\right)$ and $\left(\dot{x}_{i,q,k}^{m},\dot{y}_{i,q,k}^{m}\right)$
denote the target position target velocity, respectively. The state
of the target $q$ at the fusion time $t_{k+1}$, denoted by $\boldsymbol{s}_{t_{k+1}}^{q}$,
is given by 
\begin{equation}
\boldsymbol{s}_{t_{k+1}}^{q}=\left[x_{t_{k+1}}^{q},\dot{x}_{t_{k+1}}^{q},y_{t_{k+1}}^{q},\dot{y}_{t_{k+1}}^{q}\right]^{T}.
\end{equation}
It is assumed that the measurements are correctly associated with
their radar, i.e., no data-association uncertainty exists. Therefore,
the state transition and measurement model is
\begin{equation}
\begin{cases}
\boldsymbol{s}_{t_{k+1}}^{q}=f\left(\boldsymbol{s}_{i,q,k}^{m},t_{k+1}-t_{i,q,k}^{m}\right)+\boldsymbol{\gamma}_{k}^{q}\\
\boldsymbol{y}_{i,q,k}^{m}=h\left(\boldsymbol{s}_{i,q,k}^{m},i\right)+\boldsymbol{w}_{i,q,k}^{m},
\end{cases}
\end{equation}
where $f\left(\cdot\right)$ is the transition function with $f\left(\boldsymbol{s}_{i,q,k}^{m},t_{k+1}-t_{i,q,k}^{m}\right)=\boldsymbol{F}_{k}^{q}\boldsymbol{s}_{i,q,k}^{m}$
and
\begin{equation}
\boldsymbol{F}_{k}^{q}=\text{diag}\left(\left[\boldsymbol{I}_{2}\otimes\left[\begin{array}{cc}
1 & t_{k+1}-t_{i,q,k}^{m}\\
0 & 1
\end{array}\right]\right]\right),
\end{equation}
 and $\boldsymbol{\gamma}_{k}^{q}\sim\mathcal{N}\left(\boldsymbol{0},\boldsymbol{\Gamma}_{k}^{q}\right)$
represents the process noise, $h\left(\cdot\right)$ is the measurement
function given by
\begin{equation}
\footnotesize{h\left(\boldsymbol{s}_{i,q,k}^{m},i\right)=\left[\begin{array}{c}
R_{i,q,k}^{m}\\
\theta_{i,q,k}^{m}
\end{array}\right]=\left[\begin{array}{c}
\sqrt{\left(x_{i,q,k}^{m}-x_{i}\right)^{2}+\left(y_{i,q,k}^{m}-y_{i}\right)^{2}}\\
\arctan\left[\frac{y_{i,q,k}^{m}-y_{i}}{x_{i,q,k}^{m}-x_{i}}\right]
\end{array}\right],}
\end{equation}
and $\boldsymbol{w}_{i,q,k}^{m}\sim\mathcal{N}\left(\boldsymbol{0},\boldsymbol{\Sigma}_{i,q,k}^{m}\right)$
represents the measurement error. We further let $\boldsymbol{\Sigma}_{i,q,k}^{m}=\text{diag}\left(\sigma_{R_{i,q,k}^{m}}^{2},\sigma_{\theta_{i,q,k}^{m}}^{2}\right),$
where $\sigma_{R_{i,q,k}^{m}}^{2}$ and $\sigma_{\theta_{i,q,k}^{m}}^{2}$
are the lower bounds of the MSE error of the corresponding measures.
According to \cite{skolnik1960theoretical}, 
\begin{equation}
\begin{cases}
\sigma_{R_{i,q,k}^{m}}^{2}= & \frac{\sum_{j=1}^{J}\left|\alpha_{i,j}^{c}\right|^{2}P_{c,k}^{j}+\sigma_{r,i}^{2}}{P_{i,q,k}^{m}T_{i,q,k}^{m}}\eta_{i,q,k}^{m}\zeta_{i}^{2}c_{R}\\
\sigma_{\theta_{i,q,k}^{m}}^{2}= & \frac{\sum_{j=1}^{J}\left|\alpha_{i,j}^{c}\right|^{2}P_{c,k}^{j}+\sigma_{r,i}^{2}}{P_{i,q,k}^{m}T_{i,q,k}^{m}}\eta_{i,q,k}^{m}B_{i}^{2}c_{\theta},
\end{cases}
\end{equation}
where $\sigma_{r,i}^{2}$ is the variance of the noise, $\eta_{i,q,k}^{m}$
is the RCS of the target $q$ at the receiving time $t_{i,q,k}^{m}$,
$\zeta_{i}$ and $B_{i}$ are the transmit signal bandwidth and the
3dB receive beam width of radar $i$ , respectively, both $c_{R}$
and $c_{\theta}$ account for the unrelated constants. Therefore,
we can rewrite $\boldsymbol{\Sigma}_{i,q,k}^{m}$ as 
\begin{equation}
\boldsymbol{\Sigma}_{i,q,k}^{m}=\frac{\sum_{j=1}^{J}\left|\alpha_{i,j}^{c}\right|^{2}P_{c,k}^{j}+\sigma_{r,i}^{2}}{P_{i,q,k}^{m}T_{i,q,k}^{m}}\boldsymbol{C}_{i,q,k}^{m}\label{eq:7}
\end{equation}
where $\boldsymbol{C}_{i,q,k}^{m}$ is the constant matrix unrelated
to $P_{c,k}^{j}$, $P_{i,q,k}^{m}$ and $T_{i,q,k}^{m}$.

\subsection{Composition Measurements and the CRB}

Recall that all heterogeneous radars work in an asynchronous manner.
For target $q$ at the fusion time $t_{k+1}$, a composition measure
(CM) denoted by $\hat{\boldsymbol{s}}_{t_{k+1}}^{q}$ will be constructed
as an estimate of the true state $\boldsymbol{s}_{t_{k+1}}^{q}$.

We collect all the measurements during the $k$-th fusion interval
and define
\begin{equation}
\boldsymbol{y}_{k}^{q}=\left[\left(\boldsymbol{y}_{1,q,k}^{1}\right)^{T},\ldots,\left(\boldsymbol{y}_{1,q,k}^{M_{1,q,k}}\right)^{T},\ldots,\left(\boldsymbol{y}_{N,q,k}^{M_{N,q,k}}\right)^{T}\right]^{T},
\end{equation}
Since they are independent, the probability density function conditioned
on $\boldsymbol{s}_{t_{k+1}}^{q}$is 
\begin{equation}
p\left(\boldsymbol{y}_{k}^{q}|\boldsymbol{s}_{t_{k+1}}^{q}\right)=\prod_{i=1}^{N}\prod_{m}^{M_{i,q,k}}\mathcal{N}\left(h\left(\boldsymbol{s}_{i,q,k}^{m},i\right),\boldsymbol{\Sigma}_{i,q,k}^{m}\right),
\end{equation}
which is Gaussian with the mean $h\left(\boldsymbol{s}_{i,q,k}^{m},i\right)$
and covariance matrix $\boldsymbol{\Sigma}_{i,q,k}^{m}$. 

The CM $\hat{\boldsymbol{s}}_{t_{k+1}}^{q}$ can be constructed via
the maximum likelihood estimation (MLE) as
\begin{equation}
\hat{\boldsymbol{s}}_{t_{k+1}}^{q}=\arg\underset{\hat{\boldsymbol{s}}_{t_{k+1}}^{q}}{\max}\left[\ln\left(p\left(\boldsymbol{y}_{k}^{q}|\hat{\boldsymbol{s}}_{t_{k+1}}^{q}\right)\right)\right],\label{eq:9}
\end{equation}
for which the iterative least square method proposed in \cite{klein2016tracking}
is performed. 

For any unbiased estimator, the mean squared error of any estimator
is bounded by the CRB, i.e., 
\begin{equation}
\mathbb{E}_{\boldsymbol{y}_{k}^{q}}\left[\left(\hat{\boldsymbol{s}}_{t_{k+1}}^{q}-\boldsymbol{s}_{t_{k+1}}^{q}\right)\left(\hat{\boldsymbol{s}}_{t_{k+1}}^{q}-\boldsymbol{s}_{t_{k+1}}^{q}\right)^{T}\right]\succeq\boldsymbol{J}_{\boldsymbol{y}_{k}^{q}}^{-1}\left(\boldsymbol{s}_{t_{k+1}}^{q}\right),
\end{equation}
where $\boldsymbol{J}_{\boldsymbol{y}_{k}^{q}}\left(\hat{\boldsymbol{s}}_{t_{k+1}}^{q}\right)$
is the FIM given by 
\begin{equation}
\footnotesize{\begin{aligned} & \boldsymbol{J}_{\boldsymbol{y}_{k}^{q}}\left(\boldsymbol{s}_{t_{k+1}}^{q}\right)\\
= & \sum_{i=1}^{N}\sum_{m}^{M_{i,q,k}}\frac{P_{i,q,k}^{m}T_{i,q,k}^{m}}{\sum_{j=1}^{J}\left|\alpha_{i,j}^{c}\right|^{2}P_{c,k}^{j}+\sigma_{r,i}^{2}}\boldsymbol{H}_{i,q,k}^{mT}\left(\boldsymbol{C}_{i,q,k}^{m}\right)^{-1}\boldsymbol{H}_{i,q,k}^{m},
\end{aligned}
}
\end{equation}
where $\boldsymbol{H}_{i,q,k}^{m}=\nabla_{\boldsymbol{s}_{t_{k+1}}^{q}}h\left(\boldsymbol{s}_{i,q,k}^{m},i\right)|_{\hat{\boldsymbol{s}}_{t_{k+1}}^{q}}$.
The estimate $\hat{\boldsymbol{s}}_{t_{k+1}}^{q}$ is statistically
efficient \cite{klein2016tracking}, and thereby the CRB $\boldsymbol{J}_{\boldsymbol{y}_{k}^{q}}^{-1}\left(\boldsymbol{s}_{t_{k+1}}^{q}\right)$
is an appropriate approximate of the covariance matrix of the CM $\hat{\boldsymbol{s}}_{t_{k+1}}^{q}$. 

\section{Problem Formulation}

It is assumed that $P_{i,q,k}\triangleq P_{i,q,k}^{1}=\dots=P_{i,q,k}^{M_{i,q,k}}$
and $T_{i,q,k}\triangleq T_{i,q,k}^{1}=\dots=T_{i,q,k}^{M_{i,q,k}},\forall i=1,\ldots,N$.
The CM $\hat{\boldsymbol{s}}_{t_{k+1}}^{q}$ and its CRB matrix $\boldsymbol{J}_{\boldsymbol{y}_{k}^{q}}^{-1}\left(\boldsymbol{s}_{t_{k+1}}^{q}\right)$
serve as the measurement and corresponding covariance matrix to Kalman
filter, respectively. The filtered estimate $\tilde{\boldsymbol{s}}_{t_{k+1}}^{q}$
satisfies
\begin{equation}
\mathbb{E}_{\boldsymbol{y}_{k}^{q}}\left[\left(\tilde{\boldsymbol{s}}_{t_{k+1}}^{q}-\boldsymbol{s}_{t_{k+1}}^{q}\right)\left(\tilde{\boldsymbol{s}}_{t_{k+1}}^{q}-\boldsymbol{s}_{t_{k+1}}^{q}\right)^{T}\right]\succeq\boldsymbol{B}^{-1}\left(\boldsymbol{s}_{t_{k+1}}^{q}\right),
\end{equation}
where $\boldsymbol{B}\left(\boldsymbol{s}_{t_{k+1}}^{q}\right)$ is
the Bayesian FIM given by \cite{yan2014prior}
\begin{equation}
\footnotesize{\boldsymbol{B}\left(\boldsymbol{s}_{t_{k+1}}^{q}\right)=\sum_{i=1}^{N}\frac{P_{i,q,k}T_{i,q,k}}{\sum_{j=1}^{J}\left|\alpha_{i,j}^{c}\right|^{2}P_{c,k}^{j}+\sigma_{r,i}^{2}}\boldsymbol{D}_{i,q,k}+\tilde{\boldsymbol{B}}\left(\boldsymbol{s}_{t_{k}}^{q}\right),}\label{eq:14}
\end{equation}
where $\boldsymbol{D}_{i,q,k}=\sum_{m}^{M_{i,q,k}}\hat{\boldsymbol{H}}_{i,q,k}^{mT}\left(\boldsymbol{C}_{i,q,k}^{m}\right)^{-1}\hat{\boldsymbol{H}}_{i,q,k}^{m}$,
$\hat{\boldsymbol{H}}_{i,q,k}^{m}=\nabla_{\boldsymbol{s}_{t_{k+1}}^{q}}h\left(\boldsymbol{s}_{i,q,k}^{m},i\right)|_{\bar{\boldsymbol{s}}_{t_{k+1|k}}^{q}}$
($\bar{\boldsymbol{s}}_{t_{k+1|k}}^{q}$ is \emph{a priori} estimate
of $\boldsymbol{s}_{t_{k+1}}^{q}$ by Kalman filter), and $\tilde{\boldsymbol{B}}\left(\boldsymbol{s}_{t_{k}}^{q}\right)=\left[\boldsymbol{\Gamma}_{k}^{q}+\boldsymbol{F}_{k}^{q}\boldsymbol{B}^{-1}\left(\boldsymbol{s}_{t_{k}}^{q}\right)\boldsymbol{F}_{k}^{qT}\right]^{-1}$.
Since the diagonal elements of $\boldsymbol{B}^{-1}\left(\boldsymbol{s}_{t_{k+1}}^{q}\right)$
are heterogeneous, by defining $\boldsymbol{\Lambda}=\text{diag}\left(\boldsymbol{I}_{2}\otimes\left[\begin{array}{cc}
1 & 0\\
0 & T_{0}
\end{array}\right]\right)$, the Bayesian CRB based performance metric is given by
\begin{equation}
\footnotesize{g\left(\left\{ P_{c,k}^{j}\right\} ,\left\{ P_{i,q,k}\right\} ,\left\{ T_{i,q,k}\right\} \right)=\sum_{q=1}^{Q}\frac{1}{\text{Tr}\left(\boldsymbol{\Lambda}\boldsymbol{B}^{-1}\left(\boldsymbol{s}_{t_{k+1}}^{q}\right)\boldsymbol{\Lambda}^{T}\right)}.}
\end{equation}

For the communications, the throughput of downlink $j$ is
\begin{equation}
\begin{aligned} & r\left(P_{c,k}^{j},\left\{ P_{i,q,k}\right\} ,\left\{ T_{i,q,k}\right\} \right)\\
= & \text{\ensuremath{\log}}\left(1+\frac{P_{c,k}^{j}T_{0}}{\sum_{i=1}^{N}\sum_{q=1}^{Q}M_{i,q,k}\left|\alpha_{j,i}^{r}\right|^{2}P_{i,q,k}T_{i,q,k}+\sigma_{c}^{2}T_{0}}\right).
\end{aligned}
\end{equation}

By defining the optimization vector as 
\begin{equation}
\begin{aligned}\boldsymbol{z}_{k}= & \left[P_{1,1,k},\ldots,P_{N_{c},Q,k},T_{N_{c}+1,1,k},\ldots,T_{N_{c}+N_{p},Q,k},\right.\\
 & \left.P_{c,k}^{1},\ldots,P_{c,k}^{J}\right]^{T},
\end{aligned}
\end{equation}
the optimization problem is then formulated as
\begin{equation}
\begin{aligned} & \underset{\boldsymbol{z}_{k}}{\text{maximize}} &  & g\left(\boldsymbol{z}_{k}\right)\\
 & \text{subject to} &  & r\left(\boldsymbol{z}_{k}\right)\ge\epsilon_{k}^{j},\forall j\\
 &  &  & \sum_{q=1}^{Q}M_{i,q,k}P_{i,q,k}\le P_{total}^{i},\forall i\in\varphi_{c}\\
 &  &  & \sum_{q=1}^{Q}M_{i,q,k}T_{i,q,k}\le T_{total}^{i},\forall i\in\varphi_{p},\\
 &  &  & \sum_{j=1}^{J}P_{c,k}^{j}\le P_{total}^{c},
\end{aligned}
\label{eq:28}
\end{equation}
where the first constraint represents the throughput requirement on
the communication link, and the remaining constraints represent the
resource budgets on power and dwell time.

\section{Proposed Method for Resource Allocation}

Problem \eqref{eq:28} is nonconvex due to the objective function.
We introduce the slack variables $\left\{ \boldsymbol{V}^{q}\right\} $
and recast problem \eqref{eq:28} equivalently into, defining $\tilde{\boldsymbol{\Lambda}}=\boldsymbol{\Lambda}^{-1}$,
\begin{equation}
\begin{alignedat}{2} & \begin{aligned}\underset{\boldsymbol{z}_{k}}{\text{max}} & \underset{\left\{ \boldsymbol{V}^{q}\right\} }{\min}\end{aligned}
 &  & \sum_{q=1}^{Q}\text{Tr}\left(\left(\boldsymbol{V}^{q}\right)^{T}\tilde{\boldsymbol{\Lambda}}^{T}\boldsymbol{B}^{q}\tilde{\boldsymbol{\Lambda}}\boldsymbol{V}^{q}\right)\\
 & \text{subject to} &  & \text{Tr}\left(\boldsymbol{V}^{q}\right)=1,\forall q\\
 &  &  & r\left(\boldsymbol{z}_{k}\right)\ge\epsilon_{k}^{j},\forall j\\
 &  &  & \sum_{q=1}^{Q}M_{i,q,k}P_{i,q,k}\le P_{total}^{i},\forall i\in\varphi_{c}\\
 &  &  & \sum_{q=1}^{Q}M_{i,q,k}T_{i,q,k}\le T_{total}^{i},\forall i\in\varphi_{p},\\
 &  &  & \sum_{j=1}^{J}P_{c,k}^{j}\le P_{total}^{c}.
\end{alignedat}
\label{eq:29}
\end{equation}
Note that this is a nonconvex maximin problem, which will be solved
via the alternating descent-ascent approach \cite{razaviyayn2020nonconvex}.

For a fixed $\boldsymbol{z}_{k}$, the inner minimization problem
with respect to $\boldsymbol{V}^{q}$ has the closed form solution
given by
\begin{equation}
\boldsymbol{V}_{\ell+1}^{q}=\left(\tilde{\boldsymbol{\Lambda}}^{T}\boldsymbol{B}_{\ell}^{q}\tilde{\boldsymbol{\Lambda}}\right)^{-1}/\text{Tr}\left[\left(\tilde{\boldsymbol{\Lambda}}^{T}\boldsymbol{B}_{\ell}^{q}\tilde{\boldsymbol{\Lambda}}\right)^{-1}\right].
\end{equation}

For a given $\left\{ \boldsymbol{V}_{\ell+1}^{q}\right\} $, the outer
maximization problem is
\begin{equation}
\footnotesize{\begin{alignedat}{2} & \underset{\boldsymbol{z}_{k}}{\text{maximize}} &  & \sum_{q=1}^{Q}\text{Tr}\left[\tilde{\boldsymbol{V}}_{q}^{T}\boldsymbol{B}^{q}\tilde{\boldsymbol{V}}_{q}\right]\\
 & \text{subject to} &  & \frac{P_{c,k}^{j}T_{0}}{\sum_{i=1}^{N}\sum_{q=1}^{Q}M_{i,q,k}\left|\alpha_{j,i}^{r}\right|^{2}P_{i,q,k}T_{i,q,k}+\sigma_{c}^{2}T_{0}}\ge e^{\epsilon_{k}^{j}}-1,\forall j\\
 &  &  & \sum_{q=1}^{Q}M_{i,q,k}P_{i,q,k}\le P_{total}^{i},\forall i\in\varphi_{c}\\
 &  &  & \sum_{q=1}^{Q}M_{i,q,k}T_{i,q,k}\le T_{total}^{i},\forall i\in\varphi_{p},\\
 &  &  & \sum_{j=1}^{J}P_{c,k}^{j}\le P_{total}^{c},
\end{alignedat}
}\label{eq:31}
\end{equation}
where $\tilde{\boldsymbol{V}}_{q}=\tilde{\boldsymbol{\Lambda}}\boldsymbol{V}_{\ell+1}^{q},$
and $\boldsymbol{B}^{q}$ is defined in \eqref{eq:14}. 

The objective function of problem \eqref{eq:31} can be rewritten
as
\begin{equation}
\footnotesize{\sum_{q=1}^{Q}\text{Tr}\left[\tilde{\boldsymbol{V}}_{q}^{T}\boldsymbol{B}^{q}\tilde{\boldsymbol{V}}_{q}\right]=\sum_{q=1}^{Q}\sum_{i=1}^{N}\frac{P_{i,q,k}T_{i,q,k}\omega_{i,q,k}^{\ell}}{\sum_{j=1}^{J}\left|\alpha_{i,j}^{c}\right|^{2}P_{c,k}^{j}+\sigma_{r,i}^{2}}+const.,}
\end{equation}
where $\omega_{i,q,k}^{\ell}=\text{Tr}\left[\left(\tilde{\boldsymbol{\Lambda}}\boldsymbol{V}_{\ell+1}^{q}\right)^{T}\boldsymbol{D}_{i,q,k}\left(\tilde{\boldsymbol{\Lambda}}\boldsymbol{V}_{\ell+1}^{q}\right)\right]\ge0$,
and $const.$ represents a constant term unrelated to $\boldsymbol{z}_{k}$.
Note that $\sum_{i=1}^{N}\frac{\sum_{q=1}^{Q}\omega_{i,q}^{\ell}P_{i,q,k}T_{i,q,k}}{\sum_{j=1}^{J}\left|\alpha_{i,j}^{c}\right|^{2}P_{c,k}^{j}+\sigma_{r,i}^{2}}=\sum_{i=1}^{N}\frac{\boldsymbol{c}_{i}^{T}\boldsymbol{z}_{k}+d_{i}}{\boldsymbol{e}_{i}^{T}\boldsymbol{z}_{k}+\sigma_{r,i}^{2}}$,
and the constraints of problem \eqref{eq:31} can be rewritten as
$\boldsymbol{A}\boldsymbol{z}\le\boldsymbol{b}$, the derivation of
which is straightforward and ignored. 

Thus, problem \eqref{eq:31} is recast into 
\begin{equation}
\begin{aligned} & \underset{\boldsymbol{z}_{k}}{\text{maximize}} &  & f\left(\boldsymbol{z}\right)\triangleq\sum_{i=1}^{N}\frac{\boldsymbol{c}_{i}^{T}\boldsymbol{z}_{k}+d_{i}}{\boldsymbol{e}_{i}^{T}\boldsymbol{z}_{k}+\sigma_{r,i}^{2}}\\
 & \text{subject to} &  & \boldsymbol{A}\boldsymbol{z}_{k}\le\boldsymbol{b},\boldsymbol{z}_{k}\ge\boldsymbol{0},
\end{aligned}
\end{equation}
The update rule for $\boldsymbol{z}$ is given by
\begin{equation}
\boldsymbol{z}_{k}^{\ell+1}=\mathcal{P}_{\mathcal{Z}}\left(\boldsymbol{z}_{k}^{\ell}+\eta\nabla_{z}f\left(\boldsymbol{z}_{k}^{\ell}\right)\right).
\end{equation}
where $\nabla_{z}f\left(\boldsymbol{z}_{\ell}\right)=\sum_{i=1}^{N}\frac{\left(\boldsymbol{e}_{i}^{T}\boldsymbol{z}+\sigma_{r,i}^{2}\right)\boldsymbol{c}_{i}-\left(\boldsymbol{c}_{i}^{T}\boldsymbol{z}+d_{i}\right)\boldsymbol{e}_{i}}{\left(\boldsymbol{e}_{i}^{T}\boldsymbol{z}_{\ell}+\sigma_{r,i}^{2}\right)^{2}}$,
and $\mathcal{P}_{\mathcal{Z}}\left(\cdot\right)$ represents the
projection onto the convex set. 

The derived algorithm is summarized in Algorithm \ref{alg:adam}.

\begin{algorithm}[H]		 	
	\caption{Proposed Method for Resource Allocation}	 	
	\label{alg:adam}	 	
	\begin{algorithmic}[1]		 		
		\Require Initial allocation $\mathbf{z}_{0}$, convergence threshold $\varepsilon$, stepsize $\eta$
		\Ensure Resource allocation vector $\mathbf{z}$  	
		\Repeat
\State{Calculate $\boldsymbol{B}^{q}$}
\State{$\boldsymbol{V}_{\ell+1}^{q}=\left(\tilde{\boldsymbol{\Lambda}}^{T}\boldsymbol{B}_{\ell}^{q}\tilde{\boldsymbol{\Lambda}}\right)^{-1}/\text{Tr}\left[\left(\tilde{\boldsymbol{\Lambda}}^{T}\boldsymbol{B}_{\ell}^{q}\tilde{\boldsymbol{\Lambda}}\right)^{-1}\right]$}
\State{update $\boldsymbol{c}_{i},\boldsymbol{d}_{i}\text{ and }\boldsymbol{e}_{i}$}
\State{$\boldsymbol{z}_{k}^{\ell+1}=\mathcal{P}_{\mathcal{Z}}\left(\boldsymbol{z}_{k}^{\ell}+\eta\nabla_{z}f\left(\boldsymbol{z}_{k}^{\ell}\right)\right)$} 
		\State{$\ell\leftarrow\ell+1$}	 		
		\Until $f\left(\boldsymbol{z}_{k}^{\ell}\right)-f\left(\boldsymbol{z}_{k}^{\ell-1}\right)\le\varepsilon_{obj}$
	\end{algorithmic}	 
\end{algorithm}

\section{Simulation Results}

\begin{figure}[t]
\begin{centering}
\includegraphics[scale=0.53]{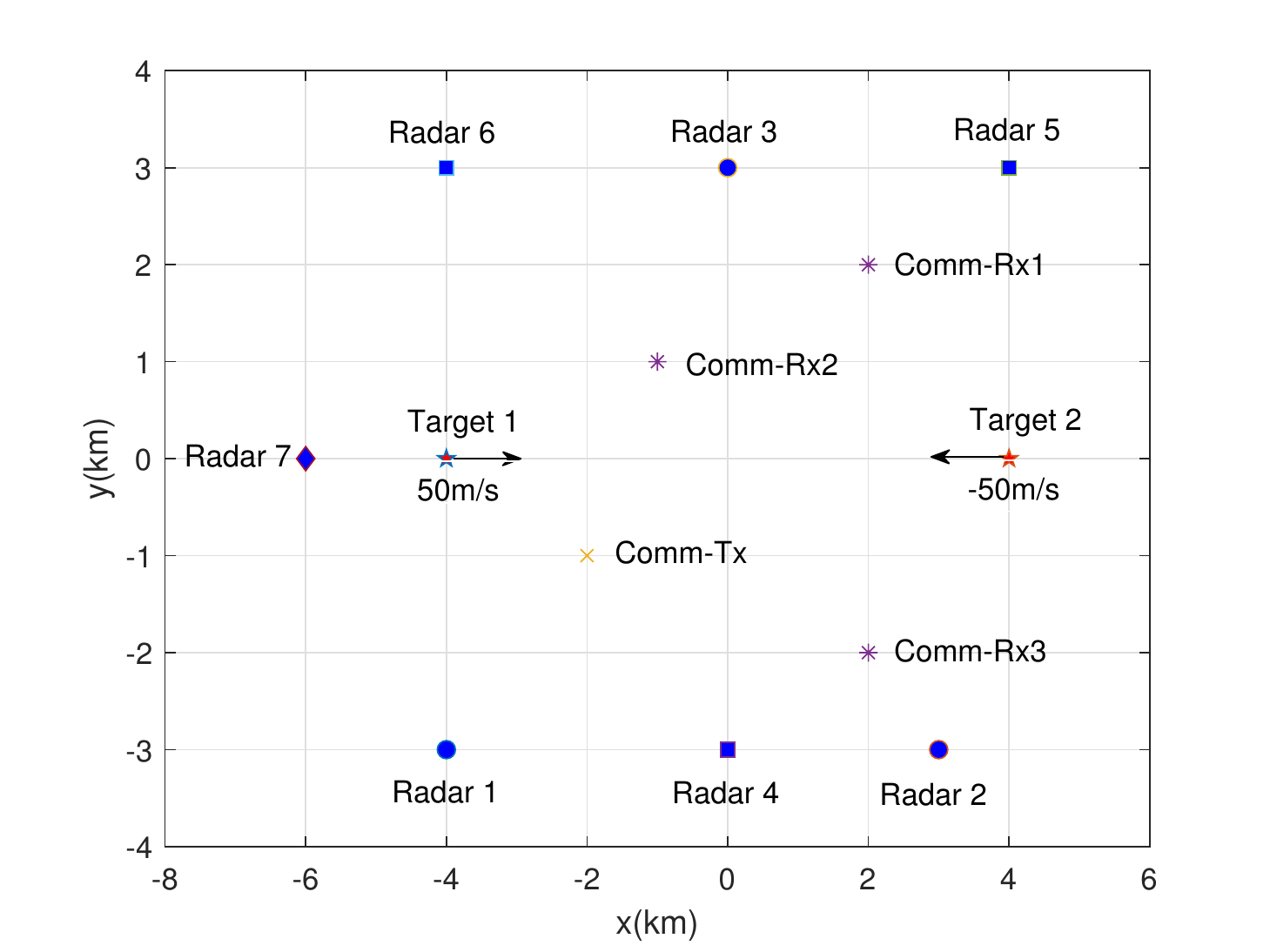}
\par\end{centering}
\caption{\label{fig:Test-scenario-of}Test scenario of the HRCN. Circle: MMR;
square: PAR; diamond: MSR}
\end{figure}

\begin{table}[t]
\caption{\label{tab:Initial-sampling-time}Initial sampling time and revisit
intervals for radars}

\centering{}%
\begin{tabular}{|c|c|c|c|c|c|c|c|c|}
\hline 
\multicolumn{2}{|c|}{Radar} & 1 & 2 & 3 & 4 & 5 & 6 & 7\tabularnewline
\hline 
\hline 
\multirow{2}{*}{Target 1} & Initial time (s) & 2 & 2.5 & 3 & 2.3 & 3.1 & 3.5 & 4\tabularnewline
\cline{2-9} \cline{3-9} \cline{4-9} \cline{5-9} \cline{6-9} \cline{7-9} \cline{8-9} \cline{9-9} 
 & revisit interval (s) & 2 & 2 & 2 & 3 & 2 & 2 & 2\tabularnewline
\hline 
\multirow{2}{*}{Target 2} & Initial time (s) & 2 & 2.5 & 3 & 2.6 & 3.2 & 3.6 & 4.2\tabularnewline
\cline{2-9} \cline{3-9} \cline{4-9} \cline{5-9} \cline{6-9} \cline{7-9} \cline{8-9} \cline{9-9} 
 & revisit interval (s) & 2 & 2 & 2 & 3 & 2 & 2 & 2\tabularnewline
\hline 
\end{tabular}
\end{table}

\begin{figure}[t]
\begin{centering}
\includegraphics[scale=0.53]{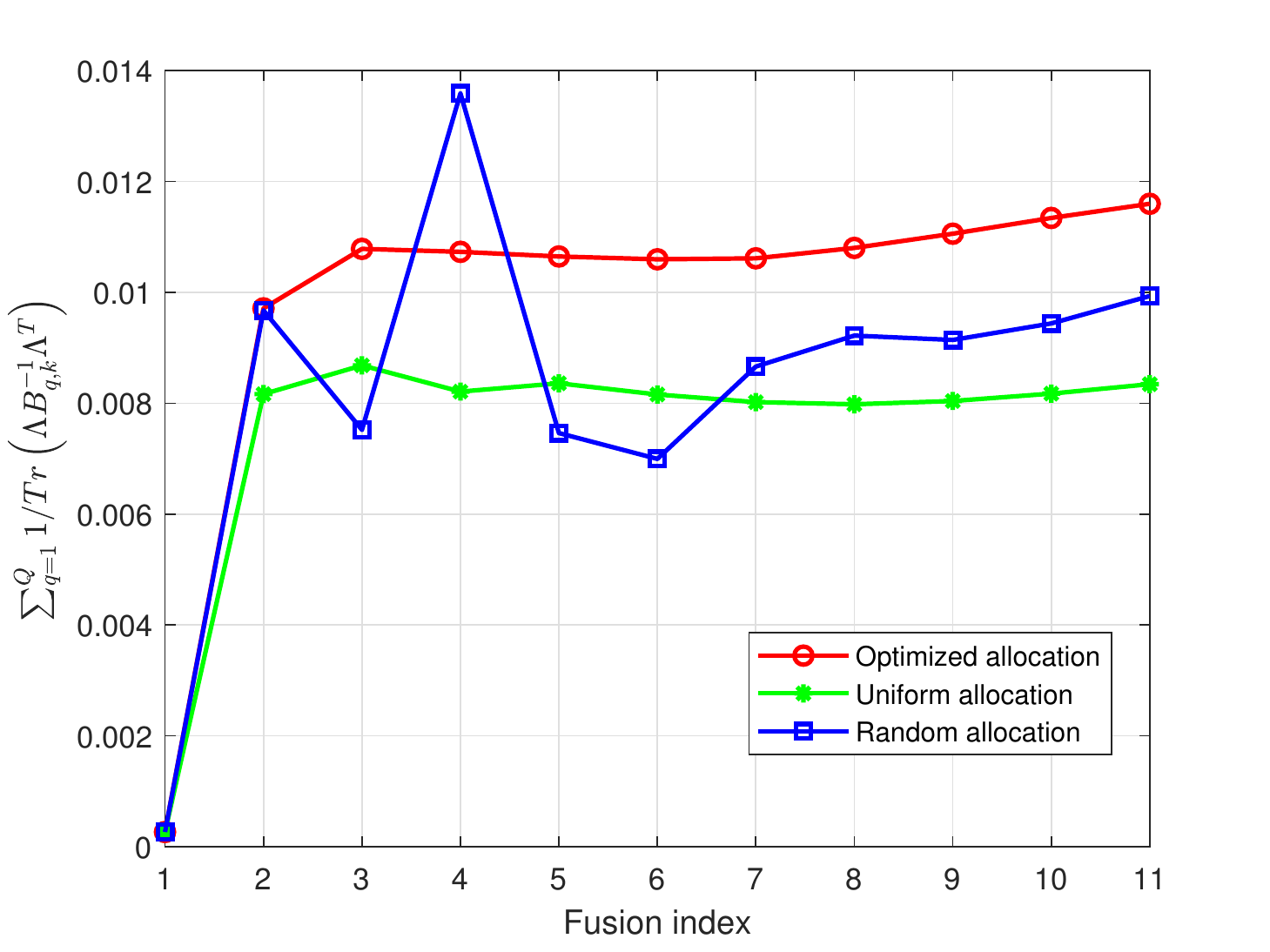}
\par\end{centering}
\caption{\label{fig:Comparison-of-the}Comparison of the Bayesian CRB based
performance.}
\end{figure}

\begin{figure}[t]
\begin{centering}
\includegraphics[scale=0.53]{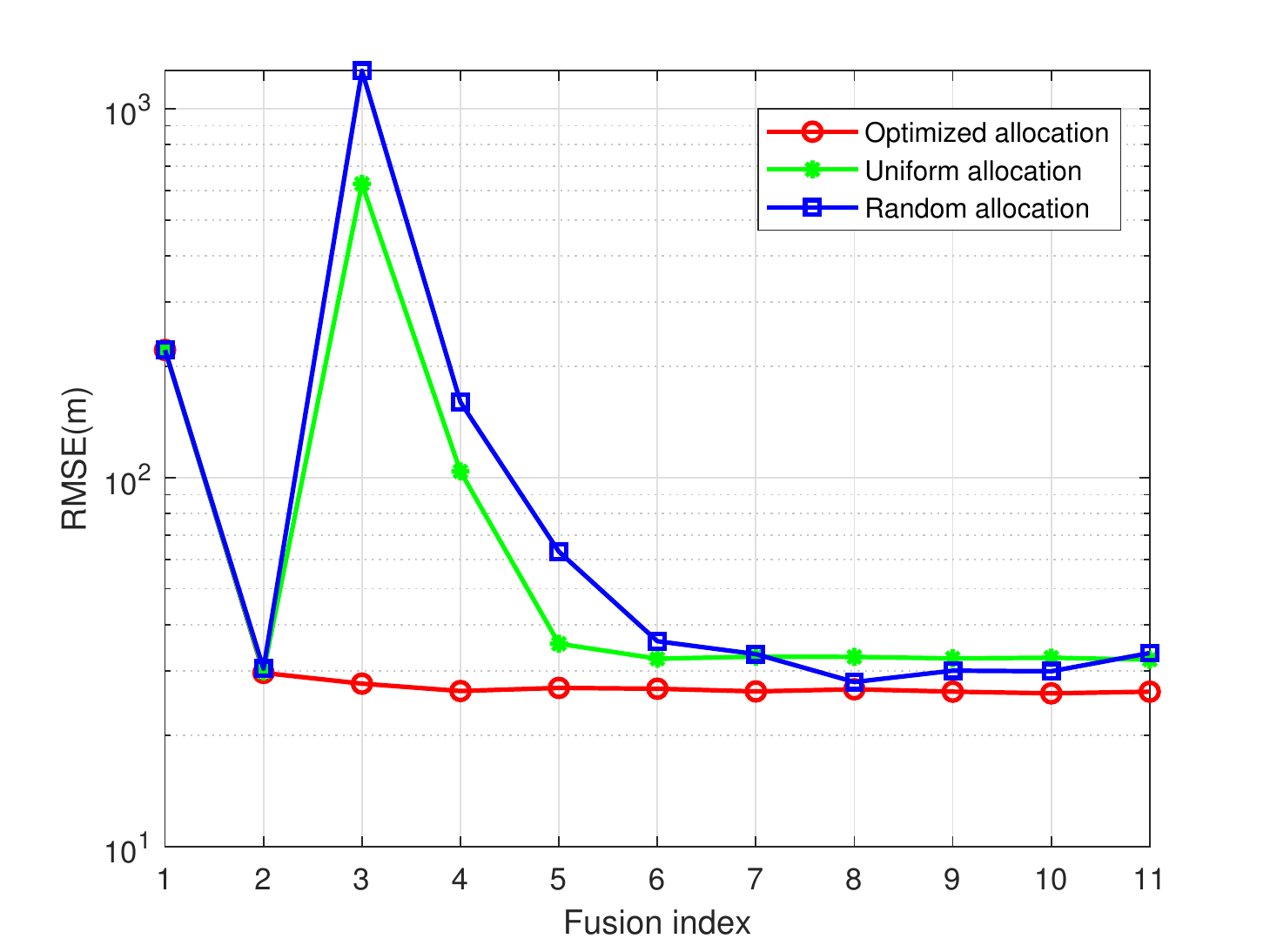}
\par\end{centering}
\caption{\label{fig:Comparison-of-the-1}Comparison of the RMSE performance.}
\end{figure}

The deployment of the HRCN is shown in Figure \ref{fig:Test-scenario-of}.
The radar system consists of $N_{cr}=3$ MMRs, $N_{pr}=2$ PARs and
$N_{mr}=1$ MSR. The communication system consists of $1$ Tx and
$3$ Rx. The radar configurations are shown in Table \ref{tab:Initial-sampling-time}.Figure
\ref{fig:Comparison-of-the} compares the achieved values of $\sum_{q=1}^{Q}1/\text{Tr}\left(\boldsymbol{\Lambda}\boldsymbol{B}_{q,k}^{-1}\boldsymbol{\Lambda}^{T}\right)$
for different allocations, where the optimized allocation is the proposed
one, the uniform allocation distributes the resource evenly, and the
random allocation is randomly generated, which might be slightly beyond
the constraints. It is clear to see that the optimized allocation
achieves the largest values over all fusion index. The tracking performances
are shown in Figure \ref{fig:Comparison-of-the-1} and Table \ref{tab:Average-RMSE-of},
where the RMSE is defined as 
\begin{equation}
\text{RMSE}=\sum_{q=1}^{Q}\sqrt{\frac{1}{N_{t}}\left(\sum_{n=1}^{N_{t}}\left\Vert \boldsymbol{\Lambda}\left(\tilde{\boldsymbol{s}}_{t_{k+1}}^{q\left(n\right)}-\boldsymbol{s}_{t_{k+1}}^{q}\right)\right\Vert ^{2}\right)},
\end{equation}
where $N_{t}$ is the number of the Monte Carlo trails. The RMSE of
the optimized allocation maintains a low level stably among all intervals
and achieves the minimum average RMSE. 

\begin{table}[t]
\caption{\label{tab:Average-RMSE-of}Comparison of the Average RMSE For the
10 fusion intervals}

\centering{}%
\begin{tabular}{|c|c|c|c|}
\hline 
Allocation & Optimized & Uniform & Random\tabularnewline
\hline 
\hline 
RMSE (m) & 26.9051 & 98.9908 & 171.2642\tabularnewline
\hline 
\end{tabular}
\end{table}

\section{Conclusions}

We consider the resource allocation for the heterogeneously-distributed
radar and communication network. The allocation aims to minimize the
Bayesian CRB based metric of target state estimation subject to some
resource budget and communication throughput requirements. The allocation
scheme designed by the proposed method shows its efficacy by the numerical
experiments.

\section*{Acknowledgment}

This work was supported by the ERC project AGNOSTIC and the FNR CORE
project SPRINGER.

\bibliographystyle{ieeetr}
\bibliography{reference,IEEEabrv}

\end{document}